
\def\figin#1{\vskip2in} 

\input harvmac

\def\s{\sigma}
\def\e{\eta}
\def\k{\kappa}
\def\ee{{\rm e}^}
\def\d{{\rm d}}
\def\IR{\relax{\rm I\kern-.18em R}}
\font\cmss=cmss10 \font\cmsss=cmss10 at 7pt
\def\IZ{\relax\ifmmode\mathchoice
{\hbox{\cmss Z\kern-.4em Z}}{\hbox{\cmss Z\kern-.4em Z}}
{\lower.9pt\hbox{\cmsss Z\kern-.4em Z}}
{\lower1.2pt\hbox{\cmsss Z\kern-.4em Z}}\else{\cmss Z\kern-.4em Z}\fi}

\def\delbar{\overline\del}
\def\inbar{\vrule height1.5ex width.4pt depth0pt}
\def\IC{\relax\thinspace\hbox{$\inbar\kern-.3em{\rm C}$}}
\def\figinsert#1#2#3{\topinsert\figin{#2}\centerline{\vbox{\baselineskip12pt
\advance\hsize by -1truein\noindent\footnotefont{\bf Fig.~\xfig#1:} #3}}
\bigskip\endinsert}

\noblackbox

\Title{\vbox{ \baselineskip12pt
\hbox{hepth@xxx/9204006}%
\hbox{NEIP92-003}
}}
{\vbox{
\centerline{
 Noncompact Coset Spaces in}
   \vskip2pt\centerline{
    String Theory}}}

\centerline{
Fernando Quevedo\footnote{$^*$}
{ Presented at BMWS workshop, HARC, Houston, January 1992.
Supported by Swiss National Foundation. email: quevedo@iph.unine.ch.}}

\bigskip
\centerline{Institut de Physique}
\centerline{Universit\'e de Neuch\^atel,}\centerline{Rue A.-L. Breguet 1
CH-2000
, Switzerland.}

\vskip .6in
\noindent A brief overview of strings propagating on
noncompact coset spaces $G/H$ is presented
in terms of gauged WZW models.
The role played by isometries in the existence of target space duality
and by
fixed points of the gauged transformations in the existence of  singularities
and horizons, is emphasized.
 A general classification of the spaces with a single
time-like coordinate is presented. The spacetime
geometry of a class of models, existing for every
dimension and having cosmological and black hole-like
interpretations, is discussed.

\Date{3/92}


\lref\bn{I. Bars and D. Nemechansky, Nucl. Phys. B348 (1991) 89.}
\lref\witten{E. Witten, Phys. Rev. D44 (1991) 314.}
\lref\rwzw{E. Witten, Comm. Math. Phys. 92 (1984) 455.\semi
For a review see P. Goddard and D. Olive, Int. J. Mod. Phys. 1 (1986) 303.}
\lref\wzwg{E. Witten, Nucl. Phys. B223 (1983) 422\semi
K. Bardacki, E. Rabinovici and B. Saering, Nucl. Phys. B301 (1988) 151 \semi
D. Karabali and H. J. Schnitzer, Nucl. Phys. B329 (1990) 649, and references
therein.}
\lref\busc{T. Buscher, Phys. Lett. 194B (1987) 59;
 Phys. Lett. 201B (1988) 466.}
\lref\vhol{For a review see, J. W. van Holten, Z. Phys. C27:57 (1985).}
\lref\gmv{M Gasperini, J. Maharana and G. Veneziano, CERN preprint
CERN-TH-6214/91 (1991).}
\lref\sen{A. Sen, preprints  TIFR/TH/91-35 and TIFR/TH/91-37\semi
S. Hassan and A. Sen, preprint TIFR/TH/91-40 (1991).}
\lref\md{M. Duff, talk at Trieste summer 1989, Nucl. Phys. B335 (1990) 610.}
\lref\muller{M. Muller, Nucl. Phys. B337 (1990) 37.}
\lref\cfmp{C. Callan, D. Friedan, E. Martinec and M. Perry,  Nucl.
Phys. B262 (1985) 593.}
\lref\kir{E.B. Kiritsis, Mod. Phys. Lett. A6 (1991) 2871.}
\lref\vero{M. Ro\v cek and E. Verlinde, IAS preprint IASSNS-HEP-91/68
(hepth@xxx/\-9110053).}
\lref\giva{P. Ginsparg and C. Vafa, Nucl. Phys. B289 (1987) 414 \semi
E. Alvarez and M. Osorio, Phys. Rev. D40 (1989) 1150.}
\lref\grvsw{A. Giveon, E. Rabinovici and G. Veneziano, Nucl. Phys. B322 (1989)
167\semi
A. Shapere and F. Wilczek, Nucl. Phys. B320 (1989) 669.}
\lref\gz{M.K. Gaillard and B. Zumino, Nucl. Phys. B193 (1981) 221.}
\lref\cfg{S. Cecotti, S. Ferrara and Girardello, Nucl. Phys. B308 (1988) 436.}
\lref\iltq{L.E. Ib\'a\~nez, D. L\"ust, F. Quevedo and S. Theisen
unpublished (1990) \semi G. Veneziano, Phys. Lett. B265 (1991) 287.}
\lref\bmq{C. Burgess, R. Myers and F. Quevedo unpublished (1991) \semi
A. Tseytlin, Mod. Phys. Lett. A6 (1991) 1721.}
\lref\bv{R. Brandenberger and C. Vafa, Nucl. Phys. B316 (1989) 391.}
\lref\mv{K. Meissner and G. Veneziano, Phys. Lett. 267B (1991) 33\semi
M. Gasperini and G. Veneziano, CERN-TH-6321/91 (hepth@xxx/\-9112044).}
\lref\tv{A. Tseytlin and C. Vafa, Harvard preprint HUTP-91/A049
(hepth@xxx/\-9109048).}
\lref\dvv{R. Dijgkraaf, E. Verlinde, and H. Verlinde, IAS preprint
IASSNS-HEP-91/22\semi
A. Giveon, Mod. Phys. Lett. A6 (1991) 2843.}
\lref\rAGG{L. Alvarez-Gaum\'e and P. Ginsparg, Ann. Phys. 161 (1985) 423.}
\lref\rpglh{P. Ginsparg, ``Applied conformal field theory,''
Les Houches lectures (summer, 1988), published in
{\it Fields, Strings, and Critical Phenomena\/},
ed.\ by E. Br\'ezin and J. Zinn-Justin, North Holland (1989).}
\lref\rvafa{C. Vafa, ``Topological Mirrors and Quantum Rings'',
Harvard preprint HUTP-91/A059 (hepth@xxx/\-9111017).}
\lref\px{P. Candelas, X.C. de la Ossa, P.S. Green and
L. Parkes, Nucl. Phys. B359 (1991) 21; Phys. Lett. 258B (1991) 118.}
\lref\barst{I. Bars, University of Southern California preprint USC-91/HEP-B4.}
\lref\hh{J.H. Horne and G. Horowitz,
University of California preprint UCSBTH-91-39 (hepth@xxx/9108001)
\semi J. Horne, G. Horowitz and A. Steif, University of California preprint
UCSBTH-91-53 (hepth@xxx/9110065);{}
 G. Horowitz, these proceedings.}
\lref\hhs{J. Horne, G. Horowitz and A. Steif, University of California preprint
UCSBTH-91-53 (hepth@xxx/9110065).}
\lref\giro{A. Giveon and M. Ro\v cek, IAS preprint IASSNS-HEP-91/84
(hepth@xxx/\-9112070).}
\lref\ind {S.P. Khastgir and A. Kumar Bubaneswar, preprint
(hepth@xxx/9109026).}
\lref\cresc{M. Crescimanno, Berkeley preprint LBL-30947\semi
P. Ho\v rava Chicago preprint EFI-91-57 (hepth@xxx/9110067)\semi
I. Bars and K. Sfetsos, Univ. Southern California preprints
USC-91/HEP-B5 (hepth@xxx/\-9110054) and USC-91/HEP-B6 (hepth@xxx/\-9111040)
\semi D. Gershon, preprint TAUP-1937-91 (hepth@xxx/9202005).}
\lref\dlp{L. Dixon, J. Lykken and M. Peskin, Nucl. Phys. B325 (1989) 325.}
\lref\barsc{I. Bars, Nucl. Phys. B334 (1990) 125.}
\lref\hel{S. Helgason, ``Differential Geometry, Lie Groups, and Symmetric
spaces'', Academic Press (1978)\semi
R. Gilmore, ``Lie Groups, Lie Algebras and Some of Their Applications'', Wiley
(1974).}
\lref\host{G. Horowitz and A. Strominger, Nucl. Phys. B360 (1991) 197.}
\lref\gique{P. Ginsparg and F. Quevedo, Los Alamos
--  Neuch\^atel preprint LA-UR-92-640, NEIP92-001
(hepth@xxx/9202092).}


 There has been a large amount of effort dedicated to find
possible connections between
String Theory and Physics. At present, the only available approach in that
direction is to study the structure and physical properties of
the semiclassical string vacua given by classes of conformal field theories
(CFT). For applications to Particle Physics, it is enough to assume
a flat four-dimensional spacetime and parametrize the different models by an
internal CFT restricted by some consistency conditions such as
worldsheet modular invariance. In this way many classes of models have been
studied and some, close to the Standard Model of Particle Physics give
hope that strings may actually be related to the real world. To ask questions
about gravitation in the context of String Theory, we have to relax the
flat spacetime assumption and substitute it by a general noncompact
CFT. These questions are of fundamental importance since the main
motivation for studying string theories is to provide a consistent way of
quantizing gravity. In particular it is of prime importance to
study singularities of cosmology and black hole-type  of
geometries in the context of String Theory, since it is  in those regimes
that the standard field theory methods of General Relativity fail to apply .
Since at the moment, coset models $G/H$ provide the most general way of
constructing CFT's, it is natural to approach these questions in terms
of noncompact  coset spaces, where the noncompactness is needed to
describe lorentzian spacetimes.

\def\tdbhtext{The causal structure of the two dimensional black hole spacetime
of \witten. Regions I,IV are asymptotic regions, regions II,III are inside
the horizon, and regions V,VI are beyond the singularities.}

The study of noncompact coset CFT's was undertaken in \dlp\
for SL(2,\IR)/$U(1)$ current algebra via the conventional
GKO construction. The formalism was later generalized to
any coset in \barsc. Given a level $k$
Kac-Moody algebra for a noncompact group $G$,
\eqn\sug{J_A(z)\, J_B(w)
\sim -{k\,\eta_{AB}/2\over (z-w)^2}+ {i\, f_{AB}{}^C\,J_C(w)
\over (z-w)}\, }
(where $g\ \eta^{AB}= f_{AC}{}^D\, f_{BD}{}^C$ is the Cartan metric and $g$ is
the Coxeter number of $G$),
the stress-energy tensor for a CFT based on $G$ is given by the Sugawara from
\eqn\set{T_G(z)={\eta^{AB}\ \colon J_A(z)\, J_B(z)\colon \over(-k+g)}\ .}
The corresponding central charge is $c_G=k\ {\rm dim}\ G/(k-g)$. For the
coset $G/H$ with stress-energy tensor $T_{G/H}=T_G- T_H$, the central
charge is $c_{G/H}=c_G-c_H$. The only changes from the compact case are
the sign of $k$ and of course the use of noncompact structure constants
$f_{AB}{}^C$. The spectrum and the
corresponding elimination of negative norm states is not yet entirely
understood for these models, and more progress is needed before we can
properly treat the string vacua obtained from this approach.

In order to have a geometrical interpretation of these vacua, we need to
have a Lagrangian formulation of these CFT's which is provided by
the WZW construction.
The WZW action for a group $G$ with
elements $g(z,\bar z)$ in complex coordinates is \rwzw,
\eqn\wzcc{L(g)={k\over{4\pi}}\int \d^2z\, \tr(g\inv\del g\, g\inv\delbar g)-
{k\over{12\pi}}\int_{B}\tr (g\inv dg\wedge g\inv dg\wedge g\inv dg)\ ,}
where the boundary of $B$ is the 2D worldsheet.
It is known that this action provides the conserved currents $g\inv\del g$
and $\delbar g g\inv$ to satisfy a chiral algebra like \sug\ and then
giving a CFT from \set\ for the group $G$.
For the coset $G/H$ \wzwg\ the standard way in nonlinear sigma models
\vhol\ is to eliminate
the $H$ degrees of freedom by gauging the
 subgroup $H$ as a symmetry of \wzcc.
To promote the global $g\to h_L\inv\, g\, h_R$ invariance to a local
$g\to h\inv_L(z)\,g\,h_R(\overline z)$ invariance, we let
$\del g\to \del g + Ag$, and $\delbar g\to \delbar g-g\bar A$. The gauge fields
transform as $A\to h_L\inv(A+\del)h_L$ and $\bar A\to h_R\inv(\bar
A+\delbar)h_R$ (so that $Dg\to h_L\inv Dg\, h_R$ for $D$ equal to either
holomorphic or anti-holomorphic covariant derivative).
Vector gauge transformations correspond to $h_L=h_R$, and axial gauge
transformations to $h_L=h_R\inv$.
Substituting covariant derivatives in \wzcc\ gives the gauged action
\eqn\wzwmcc{L(g,A)=L(g)+{k\over{2\pi}}\int \d^2z\,\tr\bigl( A\,\delbar g g\inv-
 \bar A \,g\inv\del g - g\inv A g \bar A \bigr)\ .}

Under the infinitesimal transformations $h_L\approx 1+\alpha$, $h_R=1+\beta$,
we have $\delta A=\del \alpha+[A,\alpha]$ and $\delta \bar A=\delbar
\beta+[\bar A,\beta]$. The anomalous variation of the effective action is (see
e.g.\ \rAGG\ for a review)
\eqn\evW{\delta W={k\over2\pi}(\alpha\delbar A+\beta\del\bar A)\ .}
The variation of the  counterterm $\tr A \bar A$, on the other
hand, is
\eqn\eCT{\delta(\tr A \bar A)=\tr\Bigl(-\beta\delbar A-\alpha\del\bar A
+(\alpha-\beta)\bigl[\bar A,A\bigr]\Bigr)\ .}
For the abelian case, we see that \eCT\ can compensate the variation
\evW\ for either $\alpha=\pm \beta$ since the commutator term automatically
vanishes.
Thus both vector and axial-vector gauging are allowed. In the non-abelian case,
only the vector gauging $\alpha=\beta$ is allowed.
If we change sign $\bar A\to -\bar A$ for the axial gauged case (to give $A$
and $\bar A$ the same transformation properties), then the gauged action may
be written
\eqn\wzwcc{L(g,A)=L(g)+{k\over{2\pi}}\int \d^2z\,
 \tr\bigl( A\,\delbar g g\inv\mp\bar A \,g\inv
 \del g +A \bar A \mp g\inv A g \bar A \bigr)\ ,}
where the upper and lower signs represent respectively vector ($g\to hgh\inv$)
and axial-vector ($g\to hgh$) gauging.

We now consider some naive properties of the geometry described by \wzwcc\
in the large $k$  limit.
Writing $A=A^a\sigma_a$ in terms of the generators $\sigma_a$ of $H$,
and integrating out the components $A^a$ classically gives the effective
action
\eqn\effcc{L=L(g)\pm{k\over{2\pi}}\int \d^2z\
 \tr(\s_b g\inv\del g)\, \tr(\s_a\delbar g g\inv)
 \,\Lambda_{ab}\inv\ ,}
with $\Lambda_{ab} \equiv \tr(\s_a \s_b \mp \s_a g \s_b g\inv)$. Notice that
singularities of $\Lambda$ occur at least at fixed points of the gauge
transformation $g\to h g h^{\mp1}$. This is because for infinitesimal
$h\approx 1+\alpha^a\,\sigma_a$, we see that a fixed point $g$ satisfies
$\sigma_a\,g\mp g \sigma_a=0$. Multiplying by $g\inv \sigma_b$ and taking the
trace, we see that $\Lambda=0$ at a fixed point.

{}From the transformation properties of the gauge fields and \eCT, we note
that in the case of $H$ abelian the ungauged axial or vector symmetry
remains a {\it global\/} symmetry, i.e.\ an isometry of the spacetime
geometry. In the non-abelian case, not even a global vestige of the
ungauged symmetry remains. In the abelian case, this implies that a fixed
point of the ungauged symmetry corresponds to a point with vanishing
Killing vector. For lorentzian signature, the surface carried into the fixed
point by the isometry will be a null surface (the norm of the Killing
vector is conserved), in general nonsingular and hence a horizon. We see that
fixed points of symmetry transformations generically give rise to metric
singularities when the symmetry is gauged and to horizons when ungauged. This
general property is the origin of the singularity/horizon duality of the
2D black hole of \witten. For the vector gauging, the metric can be written
$\d s^2=-\d a\,\d b/(1-ab)$, and the fixed point of the vector transformation
(the gauged symmetry) corresponds to $ab=1$, which is the singularity. The
fixed point of the axial transformation (the isometry) is $a=b=0$
indicating that the invariant surface $ab=0$ is null, and provides the
event horizon illustrated in \fig\ftdbh{\tdbhtext}. For the axial gauging, the
metric is identical (i.e.\ the geometry is self-dual) but the role of the
fixed points is exchanged,  implying the horizon/singularity duality
pointed out in \dvv.




\figinsert\ftdbh{\epsfxsize2in$$\epsfbox{2dbh.ps}$$}{\tdbhtext}

We can see that the gauged ``$G/H$'' WZW models considered here are not
the usual left or right $G/H$ coset spaces with standard coset metric, as
considered in the mathematical literature \hel\ and in standard treatments
of coset space nonlinear $\sigma$--models\vhol. This is because we gauge $g\to
hgh^{\mp1}$ type symmetries rather than $g\to gh$ or $g\to hg$, and as well
we include a Wess-Zumino term which can add a torsion piece to the metric.
Gauging the $H$ subgroup nonetheless eliminates the $H$ degrees of
freedom, and it is easily verified that the {\it signature\/} of the
resulting metric is the same as that of the standard coset metric. It is
therefore straightforward to impose the phenomenological restriction to
spaces with only a single timelike coordinate \bn. The only subtlety is
that the level $k$ appears in front of the action. Positive $k$,
in our sign conventions, results in a metric whose
compact generators correspond to timelike directions and noncompact generators
to spacelike directions. For
negative $k$ (when allowed by unitarity), the roles of
compact and noncompact generators are interchanged in the correspondence.

To classify all the coset CFT's with a single timelike coordinate we consider
first the case $k$ positive and examine the difference
$$N\equiv |G|_{\rm c} - |H|_{\rm c}\ ,$$
for all possible cosets, where $|G,H|_{\rm c}$ denote the number of compact
generators. To this end we employ the known classification \hel\ of
symmetric spaces $G/H$ (where $H$ is a maximal subgroup and $G$ is simple).
{}From this list we eliminate all cases
with $N>1$, since for a given $G$ modding out by smaller (non-maximal)
subgroups increases the value of $N$. For $N=1$, this leaves only
the case $SO(D-1,2)/SO(D-1,1)$ (\bn). For $N=0$, which corresponds to
maximal compact subgroup embeddings,
 we can identify the cases for which $H$ has a $U(1)$ factor,
$H=H'\times U(1)$ (hermitian symmetric spaces), so that $G/H'$ has an
additional compact generator, hence one timelike coordinate.
These  cases
exhaust all possibilities in which $G$ is a simple group.
For $k$ negative, we consider instead the difference
$N=|G|_{\rm nc} - |H|_{\rm nc}$ of noncompact generators, and find that the
only solution with $N=1$ is $SO(D,1)/SO(D-1,1)$.

For $G$ a product of simple groups and $U(1)$ factors, there are several
possibilities to consider:

\item{(i)\ } $G=G_1\otimes G_2\otimes G_3$ and $H=H_1\otimes H_2 \otimes H_3$
where $G_1/H_1$ has $N=1$, $G_2/H_2$  has
$N=0$ and $G_3/H_3$ is a (product of) compact coset(s).

\item{(ii)\ } $G=G'\otimes \IR$ where $G'/H$ has $N=0$.
 In this case $\IR$ provides the timelike coordinate.

\noindent
Other possibilities may be obtained by enlarging consideration from
semisimple groups $G$ to non-semisimple groups of potential relevance\gique.
We now consider the simplest class of coset models with a single
timelike coordinate and any number of spacelike coordinates
namely,  $SL(2,\IR)\otimes SO(1,1)^{D-2}\big/SO(1,1)$ models . In order to
find the metric in the large $k$ limit, we employ the standard procedure
in nonlinear $\sigma$--models \vhol,
i.e.\ find a parametrization of
the $G$ group elements, impose a unitary type gauge on the fields in the
$\sigma$--model action and then solve for the (non-propagating) $H$-gauge
fields to derive the $G/H$ worldsheet action. From that action
we can read off the corresponding background fields.
 We first
parametrize the group elements as $g={\rm diag} (g\dup_0, g\dup_1,
...g\dup_{D-2})$,
where
\eqn\gz{g\dup_0=\pmatrix{a&u\cr -v&b\cr} \qquad({\rm with}\ ab+uv=1)}
and
\eqn\ggi{g\dup_i=\pmatrix{\cosh r_i&\sinh r_i\cr \sinh r_i&\cosh r_i\cr}
 \quad{\rm with}\quad i=1, ..., D-2 \ .}
We choose the embedding such that the generator of $H=SO(1,1)$ is
$\s={\rm diag}(s_0,...s_{D-2})$
where
\eqn\gend{s_0=q_0\pmatrix{1&0\cr 0&-1\cr}\quad{\rm and}\quad s_i=q_i
  \pmatrix{0&1\cr 1&0\cr}\ ,}
with coefficients normalized to $\sum_{i=0}^{D-2} q_i ^2=1$.


Under the infinitesimal vector gauge transformations
$\delta g=\varepsilon(\s g-g \s)$, the parameters transform as $\delta
a=\delta b=\delta r_i=0$, and $\delta u=2\varepsilon q_0 u$, $\delta
v=-2\varepsilon q_0 v$. The choices $u=\pm v$ thus fix the gauge
completely. From $\pm u^2=1-ab$, we are left with the parameters $a, b,
r_i$ as the $D$ spacetime coordinates. Substituting  into
\effcc, we find the action (for both gauge choices $u=\pm v$):
\eqn\effd{\eqalign{L={k_0\over{2\pi}}\int \d^2z\,
 \biggl(-{\del a\delbar b +\del b\delbar a\over 2(1-ab)}
 +\sum_i \kappa_i\Bigl(\delta_{ij}+{\kappa_j\eta_i
 \eta_j\over 1-ab}\Bigr)\del r_i\delbar r_j\cr
   +{\kappa_i\e_i\over 2(1-ab)}\Bigl((b\del a-a\del b)\delbar r_i
 +\del r_i (b\delbar a-a\delbar b)\Bigr)\biggr)\cr}}
where $\k_i\equiv k_i/k_0$ and $\e_i \equiv q_i/q_0$.

This action can be identified with a $\sigma$--model action of the form
\eqn\efft{S=\int \d^2z\,\bigl(G_{MN}+B_{MN}\bigr)\,\del X^M \delbar X^N}
to read off the background metric and antisymmetric tensor field (torsion).
We see that \effd\ gives for $D=2$ the (dual) black hole metric of
\witten\ $\bigl(\d s^2={-\d a\,\d b/(1-ab)}\bigr)$.
For $\k_i\to 0$, it reduces as expected to the 2D black hole and for
$\e_i \to 0$ gives the 2D black hole times $D-2$ flat coordinates, again
as expected since in this limit $H=SO(1,1)$ is completely embedded in
$SL(2,\IR)$. Note that for any $D$ there is no torsion, in particular the WZ
term can be seen to be a total derivative for our choice of gauge.
Furthermore we can observe that there are at least $D-2$ isometries since
the metric does not depend explicitly on the coordinates $r_i$. Finally,
note that the metric blows up only at the fixed point $ab=1$ which is the
point where $\Lambda$ in \effcc\ vanishes and the classical integration is not
justified. The fixed point of the isometry $g\to hgh$ is at $ab=0$, which
we expect to lead to a horizon.

To further analyze this metric, we change to coordinates in
which it is diagonal . We consider (as in the 2D case) the regions
bounded by the horizon and singularity (\ftdbh):
\eqn\eregs{(i)\  0<ab<1\,,\quad (ii)\  ab<0\,,
\quad (iii)\ ab>1\ .}
$(i)$ corresponds to the interior regions II, III;
$(ii)$ to the asymptotic regions I, IV ;
and $(iii)$ to the additional regions V,VI.

In the interior regions $(i)$,
we can change to coordinates  $t$, $X_0$, $X_i$ by defining
\eqn\chvi{\eqalign{a&=\sin t\ \ee{(X_0+m X_{D-2})}\cr
    b&=\sin t\ \ee{-(X_0+mX_{D-2})}\cr
    r_i&=N_{ij}\, X_j\ ,\cr}}
where the coefficients $N_{ij}$ and $m$ are given in \gique.
In these coordinates the metric takes the diagonal form
\eqn\metru{\d s^2=
{k_0\over {2\pi}}\Bigl(-\d t^2+\tan^2 t\, \d X_0^2+
    \sum_{i=1}^{D-2} \d X_i^2\Bigr)\ .}

Notice that this metric is a particular case of the cosmological
solutions found in \muller.
The remaining regions are described similarly. For the asymptotic regions
$(ii)$, we find
\eqn\metrd{\d s^2
={k_0\over {2\pi}}\Bigl(\d R^2-\tanh^{2}R\, \d X_0^2
+\sum_{i=1}^{D-2} \d X_i^2\Bigr)\ .}
Finally, in the regions $(iii)$ beyond the
singularity
the metric is
\eqn\metrt{\d s^2=
{k_0\over {2\pi}}\Bigl(\d R^2-\coth^{2}R\, \d X_0^2+
    \sum_{i=1}^{D-2} \d X_i^2\Bigr)\ .}

Using the symmetry $a\to -a$, $b\to -b$, we identify the geometry (2D
black hole)$\otimes \IR^{D-2}$. In particular the isometry generated by
$g\to hgh$ is now explicit (it is a linear combination of translation in
$X_0$ and the $X_i$'s).   
We can see how the associated Killing vector changes signature on each
boundary:
 it is timelike in
\metrd\ and \metrt\ and spacelike in the region \metru\ in between.
In \effd, this was not explicit in the $a, b, r_i$ coordinates. Although we
have chosen
a general embedding of $H=SO(1,1)$ in all of $G$, the resulting
geometry nonetheless coincides with the
case $\eta_i=0$, where $SO(1,1)$ was embedded only in $SL(2,\IR)$. This is as
expected since the $SO(1,1)$ factors in $G$ are
 abelian and therefore transform trivially under $g\to hgh\inv$.
The spacetime diagram for the relevant 2D geometry  was shown in \ftdbh.
%
%


We now consider the axial gauging .
The $D=3$ case has also been discussed in \hh.
Under the infinitesimal gauge transformation
$\delta g=\varepsilon (\sigma g+g\sigma)$,
we see that $\delta u =\delta v =0$ and $\delta a=2\varepsilon q_0 a$,
$\delta b=-2\varepsilon q_0 b$, $\delta r_i=2\varepsilon q_i$.
A simple choice that fixes the gauge completely is $a=\pm b$. Using $\pm
a^2=1-uv$ leaves $u$, $v$, and $r_i$ as the spacetime coordinates.
Substituting into \effcc, and using the above gauge fixing
gives the effective action
\eqn\effs{\eqalign{L={k_0\over{2\pi}}\int
&\d^2z\,\biggl( \Bigl(\kappa_i\,\delta_{ij}
-{\kappa_i\,\kappa_j\,\eta_i\,\eta_j\over{1-uv+\rho}}\Bigr)
\del r_i\delbar r_j + {(u\del v-v\del u)
(u\delbar v-v\delbar u)\over 4(1-uv+\rho)}\cr
&\qquad\qquad -\half(\del u\delbar v + \del v\delbar u)
-{(u\del v+v\del u) (u\delbar v+v \delbar u)\over 4(1-uv)}\cr
&\qquad-{\kappa_i\,\eta_i\over 2(1-uv+\rho)}
\Bigl[(u\del v-v\del u)\delbar r_i - \del r_i (u\delbar v-v\delbar u)\Bigr]
\biggr)\ .}}
Where $\rho^2\equiv\sum{\k_i\eta_i^2}$.
{}From this expression we can make the following observations. First,
unlike the vector gauging, there is nonvanishing torsion given by the term
in square brackets in \effs, even though the WZ term vanishes for the
gauge choice made. We also can see that the metric has singularities at
$uv=1$, which in 2D is the fixed point of the axial transformation, and also at
$uv=1+\rho$, which is not a fixed point. Again the lines $uv=0$
represent horizons, and the metric and torsion do not depend
on the $r_i$ variables so there are also the $D-2$ isometries $r_i\to r_i
+ {\rm constant}$. As in the vector case, the $D=2$ ($\kappa_i = 0$) limit
reproduces the 2D black hole of \witten. Furthermore the $\eta_i=0$
limit gives the geometry (2D black hole)$\otimes \IR^{D-2}$
(with vanishing torsion) as in the
vector case, recovering the self-duality of those solutions.

\def\tdbstext{A two dimensional slice of the three dimensional black string
metric \effs. In addition to the regions of \ftdbh, the regions VII,VIII
lie between the singularities and inner horizons.}

The general case is more conveniently studied via
variables that diagonalize the metric in different regions. We will
consider the analog of the three regions $(i),(ii),(iii)$
of the vector case \eregs, but with $a,b$ exchanged for $u,v$.


It is straightforward to see that the same changes of variables
made for the vector case also diagonalize this metric, but now for $m=0$.
For $(i)$ $0<uv<1$, we find
\eqn\metruu{\d s^2=
{k_0\over{2\pi}}\Bigl(-\d t^2+{1\over{(\rho^2+1) \tan^2 t +\rho^2}}
\bigl(\d X_0^2+\tan^2t\, \d X_{D-2}^2\bigr)+\sum_{l=1}^{D-3} \d X_l^2\Bigr)\ ,}
and the antisymmetric tensor is
\eqn\torsu{B_{X_0 X_{D-2}}=\bigl((\rho^2+1)\tan^2 t +\rho^2\bigr)\inv\ .}
With similar expressions for the other regions
obtained from analytic continuations as in
\metru\ . Again,\metruu\ and \torsu\ admit a cosmological interpretation.
%

{}From these metrics we can compute the corresponding curvature scalar
in each of the regions and find
\eqn\curv{R=B_{X_0 X_{D-2}} + {\rm constant}\ .}
We see that $R$ blows up only in region
$uv>1$ at the hyperbola $uv=1+\rho^2$ which is the real singularity,
whereas the surface $uv=1$ is only a metric singularity where the
signature of the metric changes. The latter is another horizon,  in addition to
$uv=0$. The geometry is thus (3D black string)$\otimes \IR^{D-3}$
with nonvanishing torsion and an inner horizon. The 2D representation
($uv$ diagram) with the eight different regions separated
by the horizons and singularity, is presented
in \fig\ftdbs{\tdbstext}.
It is not surprising that there is a trivial $\IR^{D-3}$
crossed on since the $hgh$ action of $SO(1,1)$ only acts on one nontrivial
linear combination of $SO(1,1)$ generators of $G$.
The expression for the dilaton can be found  by considering the correct measure
in the
path integral, but it is technically simpler to find it by solving the
background field equations to lowest order in $\alpha'$. The exact
expressions can be found in \gique.

\figinsert\ftdbs{\epsfxsize2in$$\epsfbox{2dbs.ps}$$}{\tdbstext}


The two different spacetime geometries, corresponding to the vector and
axial gaugings of the $G/H$ WZW model, can be viewed as different modular
invariant combinations of representations of the same holomorphic and
anti-holomorphic chiral algebras. There are general arguments \kir\ that
show that the vector and axial gaugings are dual, in the sense of having
equal partition functions. The duality is similar to the familiar $r\to
1/r$ duality in $c=1$ conformal field theory where two seemingly different
theories are as well related by a changing the sign of the left (or
holomorphic) currents $J=J_L$ with respect to the right (or
anti-holomorphic) currents $\overline J=J_R$. The lorentzian $D=2$ case is
special since the same geometry (2D black hole) is obtained by either the
vector or axial gauging, so we say that the model is self-dual. We now
point out the sense in which geometries for $D\geq 2$ are dual, placing
the vector/axial duality in a more generalized context.

In ref.~\busc, the $r\to 1/r$ duality of compactified string theories
was generalized to any string background for which the worldsheet action
has at least one isometry. The worldsheet action for the bosonic string \effcc\
in a background with $N$ commuting isometries, can be written as
\eqn\sigmad{\eqalign{S = {1\over{4\pi\alpha'}}\int \d^2z\,
\Bigl(Q_{\mu\nu}(X_{\alpha})\,\del X^{\mu} \delbar X^{\nu}
 + Q_{\mu n}(X_{\alpha})\del X^{\mu} \delbar
X^n  + Q_{n \mu}(X_\alpha)\del X^n \delbar X^{\mu}
 \cr \qquad + Q_{mn}(X_\alpha)\del X^m \delbar X^n +
  \alpha' R^{(2)}\Phi(X_\alpha)\Bigr)\ ,\cr}}
where $Q_{MN}\equiv G_{MN}+B_{MN}$ and lower case latin indices $m,n$
label the isometry directions. Since the Lagrangian \sigmad\ depends on $X_m$
only through their derivatives, we can describe it in terms of the first order
variables $V^m$ and add an extra term to the action
$\widehat X_m(\del \overline V^m - \delbar V^m)$ imposing the constraint
$V^m=\del X^m$.
Integrating the Lagrange multipliers $\widehat X_m$ in the above
gives back \sigmad. After partial integration and solving for
$V^m$ and $\overline V^m$, we find the dual action $S'$
which has the identical form as $S$ with
the dual backgrounds  given in terms of the original ones by
\eqn\qp{\eqalign{Q'_{mn}& = Q\inv_{mn}\cr
  Q'_{\mu \nu}& = Q_{\mu \nu} - Q\inv_{mn}\,Q_{n\nu}\,Q_{\mu m}\cr
  Q'_{n \mu}& = Q\inv_{nm}\, Q_{m \mu}\cr
  Q'_{\mu n}& = -Q\inv_{mn}\, Q_{\mu m}\ .\cr}}
To preserve conformal invariance, it can be seen  \refs{\busc,
\giva} that the dilaton field has to transform as
$\Phi' = \Phi - \log\sqrt{\det G_{mn} }$
Notice that equations \qp\ reduce to the usual duality transformations for
the toroidal compactifications of \grvsw\
in the limit $Q_{m\mu}=Q_{\mu m}=0$. For the case of a single isometry
($m=n=0$), we recover the
explicit expressions of \busc. This is to our knowledge the most general
statement of duality in string theory . In particular we see that a
space with no torsion ($Q_{m\mu} =Q_{\mu m}$) can be dual to a space with
torsion ($Q'_{m\mu}=-Q'_{\mu m}$). Notice that for every isometry
we do not have to go to the first order formalism, i.e.\ we can
integrate the Lagrange multipliers $\widehat X_m$ for some of the
fields and instead the $V^m$ for the remaining fields with isometries.
This is the most general form of these duality transformations, and
eq.~\qp\  should be read with indices $m,n$ running over only
the variables with isometries that have been dualized. The
total duality group includes these transformations as well as shifts
in the antisymmetric tensor field and has been argued \cfg, \md\ to be
equivalent to $SO(N,N,\IZ)$.

 In string theory, duality symmetry was
originally discovered in toroidal compactifications and found
to interchange winding and momentum  states in the
compactified theory. We have seen that the toroidal compactification
is a particular case of a $\sigma$--model with isometries and thus
has this symmetry manifest. The interchange
of winding and momenta states realizes the duality symmetry in this
particular background but is not necessarily a generic feature of
duality, so we might expect duality even in backgrounds where
winding modes are not present. A particular example is given by the
2D Lorentzian black hole .


In order to make a connection between duality in this formulation and
the vector--axial duality
in $G/H$ WZW models, we just have to identify the correct action
of the vector (axial) isometry when gauging the axial (vector) transformation
and apply \qp\ (see also \hh.)  It is straightforward
to see that regions
II  of both geometries are interchanged. Region
V of \ftdbh\ is mapped to region I of \ftdbs\
and in particular the singularity of the first is mapped to one horizon
in the second. Also, region I of the vector gauging black hole gets mapped
to regions V and VII together of the axial gauging black hole. This has
the interesting implication that a surface which in one geometry is perfectly
regular ($ab=\rho^2$) is mapped to the singularity in the
other geometry ($uv=1+\rho^2$). This goes even further than the black hole
singularity/horizon duality of the 2D black holes \dvv, since in that case
the horizon is a better behaved region than the singularity but
there remains nontrivial behavior such as the exchange of spacelike and
timelike coordinates. In the present case it can be seen explicitly that string
theory can deal with spacetimes that have singularities at the classical level,
in the sense that there still exists a description of interactions, etc.\ for
that region of spacetime by going to the dual geometry!
\nobreak
\listrefs


\bye